# Open-Source Optimization of Hybrid Monte-Carlo Methods for Fast Response Modeling of NaI(Tl) and HPGe Gamma Detectors

Matthew Niichel*, Stylianos Chatzidakis*

*School of Nuclear Engineering, Purdue University, West Lafayette, IN 47906, mniichel@purdue.edu

**Abstract**

Modeling the response of gamma detectors has long been a challenge within the nuclear community. Significant research has been conducted to digitally replicate instruments that can cost over $100,000 and are difficult to operate outside a laboratory setting. Subsequently, there have been multiple attempts to create codes that replicate the response of sodium-iodide and high purity germanium detectors for the purpose of deriving data related to gamma ray interaction with matter. While robust programs do exist, they are often subject to export controls and/or they are not intuitive to use. Through the use of the Hybrid Monte-Carlo methods, MATLAB can be used to produce a fast first-order response of various gamma ray detectors. The combination of a graphics user interface with a numerical based script allows for an open-source and intuitive code. When benchmarked with experimental data from Co-60, Cs-137, and Na-22 the code can numerically calculate a response comparable to experimental and industry standard response codes. Through this code, it is shown that a savings in computational requirements and the inclusion of an intuitive user experience does not heavily compromise data when compared to other standard codes or experimental results. When the application is installed on a computer with 16 cores, the average time to simulate the benchmarked isotopes is 0.26 seconds and 1.63 seconds on a four-core machine. The results indicate that simple gamma detectors can be modeled in an open-source format. The anticipation for the MATLAB application is to be a tool that can be easily accessible and provide datasets for use in an academic setting requiring the gamma ray detectors. Ultimately, providing evidence that Hybrid Monte-Carlo codes in an open-source format can benefit the nuclear community.

## 1.) Introduction

Since the 1970s, non-proliferation efforts have been the corner stone of the nuclear community. A significant portion of these efforts have been dedicated to the detection of gamma ray radiation and isotope identification of radioactive materials. Since the 1970s significant progress has been made on detector technology. Advanced materials and algorithms have improved the energy resolution of most detectors to the point where identification of isotopes can be made with a high level of accuracy. However, like most of the nuclear community, detection still remains a costly endeavor. As a result, significant research has been conducted with an attempt to successfully replicate gamma detector response for the use of academia and the production of large data sets.

The United States Department of Energy and Oakridge National Laboratory maintain a repository of transport codes which have long been regarded as the industry standard for modeling the transport of gammas and neutrons. Among the long list of specialized codes, two stand out in particular for modeling detector response. The well-known Monte-Carlo Neutron Particle Transport (MCNP) can be used to track gammas and neutrons throughout a user defined universe of various materials and geometries. The second, Gamma Detector Response and Analysis Software (GADRAS) aims to replicate both gamma and neutron detectors based on user defined scintillating material and instrument parameters. While both codes provide high fidelity results when compared to experimental data, there are three significant issues that arise when making an attempt to use these programs to build models.

A combination of export controls and under manning of the Oakridge website creates a lead time of several months for requesting software. Additionally, the export controls prevent some users from applying at all due to their country of nationality. For many settings, waiting months to receive a copy of the codes creates barriers to making full use of the software. Despite the usefulness and quality of the software, the lack of timely distribution complicates wide spread use.

Additionally, user friendliness is a characteristic that is lacking with both MCNP and GADRAS. MCNP is a terminal-based program which, takes a considerable amount of time to learn. For a user that is not well versed with the windows command terminal, MCNP can be difficult to utilize to the full potential. Moreover, the input files require a highly specialized language that take additional time to learn. Although, there is online documentation and user manuals, modeling a detector's response may offer additional challenges to users over using an actual detector. Conversely, GADRAS makes use of a graphics user interface, but lacks the robust documentation that exists for MCNP. To appropriately model a detector requires trial and error that may be a deterrent for some users.

The final issue that is present with the codes, particularly with MCNP, is the computational requirements for tracking particles through the traditional Monte-Carlo methods. As a result, simple simulations can take hours or days to produce useful results. In a worst-case scenario, the computing requirement can be beyond that of which common computers can handle. The solution of which requires super-computers that are not practical for most applications.

While these three issues are not all inclusive to every user, they present a challenge which reduces the availability of modeling gamma detector response. As a result, they are better off making use of the costly detectors. Recently, research has been conducted with hybrid Monte-Carlo methods. There is strong evidence supporting reduced computational requirements to achieve similar results when compared to traditional transport code. Although the hybrid methods most likely cannot replace traditional Monte-Carlo, making use of it in detector modeling may eliminate computational limitations. Moreover, an open-source code which creates a clean user interface with intuitive inputs eliminates the export control problem and the need to learn complex coding for simple applications.

Albeit, traditional codes will remain the industry standard for the for the foreseeable future. An open-source code which implements a hybrid Monte-Carlo approach along with an intuitive interface will minimize the problems that commonly arise when modeling gamma detectors. The following is a description of a MATLAB application integrating these solutions into a fist-order approximation of thallium dopped sodium-iodide (NaI(Tl)) and high purity germanium (HPGe) detectors.

**2.) Background**

Previous work has been conducted in this area with an attempt to model Compton Scattering with Python language, while useful this model fails to address all the interactions that occur within detectors. (B. Kessler, 2023) An additional project addresses the use of parametric equations to model NaI response. However, this model makes use of MCNP for the production of models. (Neeraj, 2013)

A simulation that removes dependance on MCNP and represents a majority of radiation-matter interactions is ideal. While work has been conducted on neutron detectors, the team focuses on highly specific parameters of $BF_3$ detectors to other detector types. (Rubina , Aziz, Sikander, & Mirza, 2018) A detector suite that offers simulations for three systems and two common radiation particle interactions will offer a more comprehensive contribution to radiation modeling.

Radiation effects with matter can be treated deterministically in some specific applications. A common example is dose-rate and exposure. Deterministic methods are often conservative, because they return the same results for a given set of inputs. Due to use in safety protocols deterministic results are sought after. However, radiation is highly probabilistic and most applications require a different approach.

Traditionally, this is conducted *via* direct Monte-Carlo modeling which applies randomness to a system by statistical sampling over time. (Anderson, 1986) Direct Monte-Carlo methods require significant computational power and are a generalized approach to randomness, but the Hybrid Monte-Carlo method was created to apply random sampling with reduced computational requirements for complex modeling problems. While the decrease in computational requirement varies from problem and computer used, estimates are around 10 times faster for Hybrid Monte-Carlo methods. (Duane, Kennedy, Pendleton, & Roweth, 1987)

An attempt to utilize hybrid Monte-Carlo methods for calculating the counting efficiency of a NaI detector was conducted in 2007. The team integrated analytical relationships with traditional Monte-Carlo to simulate point-source and planer radiation fields in a 3"x3" NaI detector. Their findings support that this approach produces efficiencies similar to results produced by earlier accepted models for three discrete energies of Cs-137, Co-60, and Na-22. The overall result is that the time to compute was 0.2-0.8 seconds on a windows XP program. The results of their findings highlight the computational advantage of integrating analytical relationships with traditional Monte-Carlo methods. (Yalcin, Gurler, Kaynak, & Gundogdu, 2007)

## 2a.) Overview of Hybrid Monte-Carlo Method

The hybrid Monte-Carlo is a means of methodically sampling random distributions. For instance, Compton scattering can be efficiently modeled because the input energies are known and the approximate gamma interactions adhere to mathematical relations based on the density of the material. While scattering angles are still computationally random, the distribution of each scattering is statistically determined based on incoming energy. The pre-allocated distribution of scattering angles allows for a reduced computational requirement because it is calculated independent of each time dependent event. (Taboga, 2021) The process follows a 5-step process illustrated in figure 1. (Niichel & Chatzidakis, 2024)

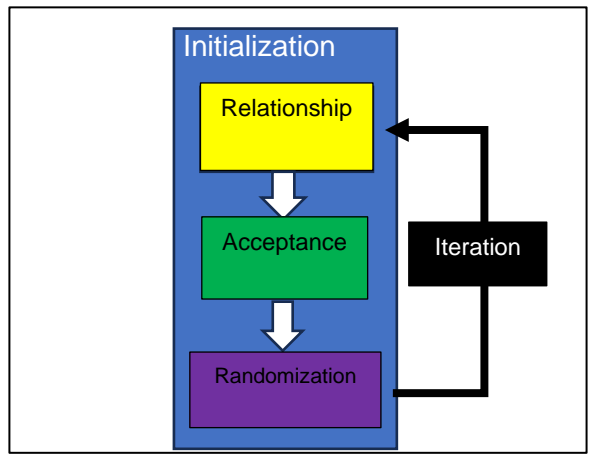

**Figure 1**. Block diagram of the hybrid Monte-Carlo Method.

The distribution is generated in the initialization block. The empirical correlations of gamma interactions are applied in the relationship block. The algorithm then conditionally accepts or rejects the outcome of inputs. Finally, randomness is applied to the inputs based on the initialized distribution. Iteration is applied to represent each event in a sample.

## 2b.) Gamma Detection Simulation Procedure

The NaI and the HPGe detectors features systems capable of sensing gamma and high energy x-ray radiation. While there are several different detector materials on the market, we focus on the two of the most readily available. The simulation procedure is show in the following block diagram in figure 2. It consists of 5 "If statements" which compare the energies of the input gammas to MATLAB generated random numbers. A version of the pseudo code is provided in appendix a.

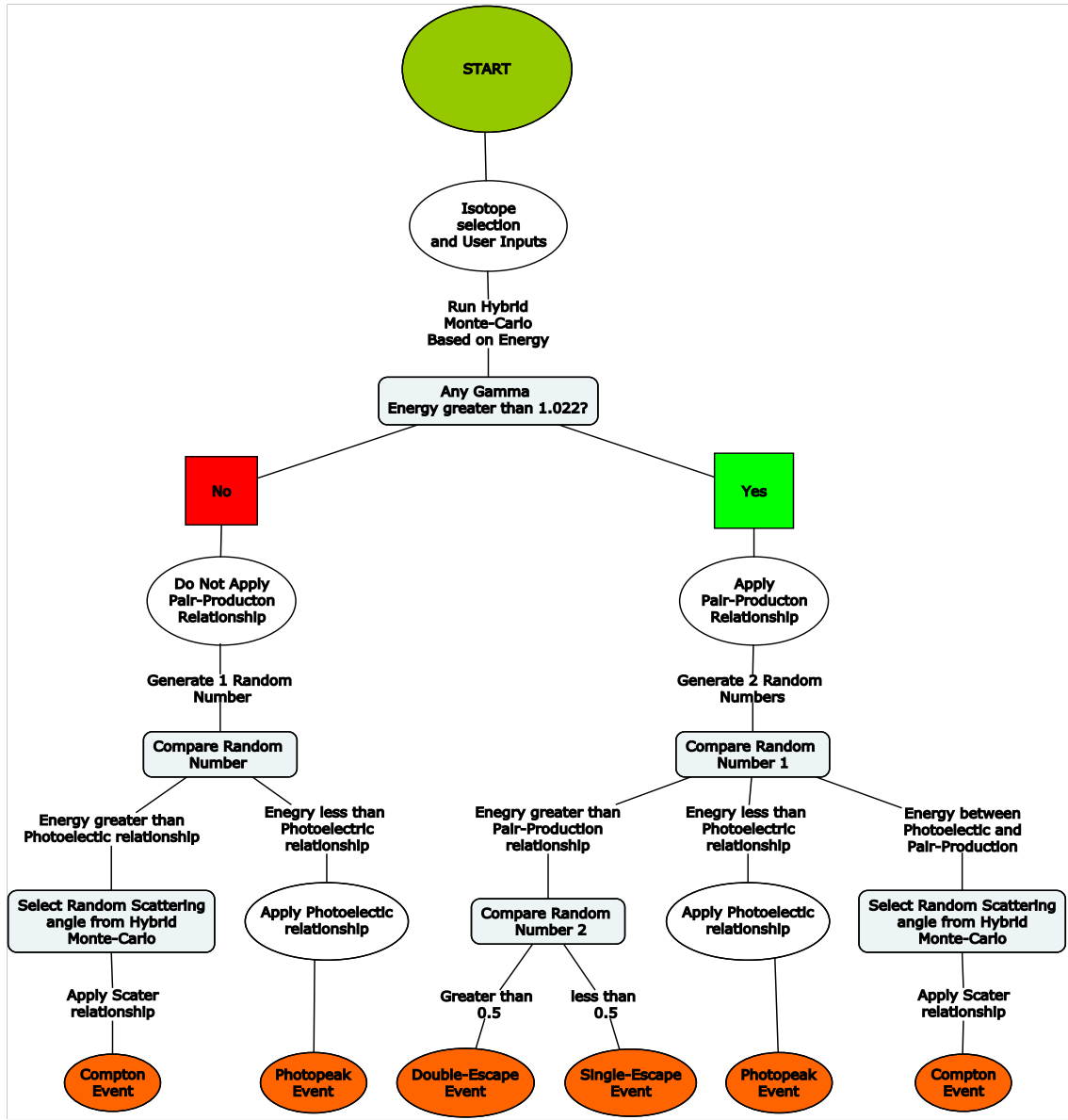

**Figure 2**. Block diagram of the algorithm comparator steps leading to gamma-matter interaction events.

The gamma photon-detector interactions are governed by three gamma interactions with matter: photoelectric effect, Compton scattering, and pair-production. These physical interactions determine the response of the detector and the output that is displayed in the graphical user interface. The fraction of energy distribution is based on the probability that all photons interacting with matter inside the detector undergo one of the aforementioned phenomena. The equation (1) is the total probability of interaction for any single source. (Knoll, 2010) The derivation for equations (2) through (4) were produced by digitizing a plot of energy dependence of the various gamma-ray interactions processes in sodium-iodide, then using MATLAB curve fitting to produce correlations. (Evans, 1955)

$$\mu_T = \mu_{pe} + \mu_{cs} + \mu_{pp} \tag{1}$$

$$\mu_{pe} = (0.000243 + 0.72010 * e^{(-7.36405*E)}) \tag{2}$$

$$\mu_{cs} = (0.0319 + 0.09681 * e^{(-1.2035*E)}) \tag{3}$$

$$\mu_{pp} = (0.000595 * e^{(0.597*E)}) \tag{4}$$

Where $\mu_T$ is the total probability of gamma interaction with matter, $\mu_{pe}$ is the probability for photoelectric effect, $\mu_{cs}$ for Compton scattering, and $\mu_{pp}$ for pair production. The values were estimated by the digitization and curve fitting using energy dependent relations for sodium-iodide. (Evans, 1955) (Roth, Primbsch, & Lin, 1984). While a good first approximation for the model, the values for photoelectric effect and Compton scattering required modification to better match benchmarked data. A comparison of the NaI averaged values shown in equations (2), (3) and (4) and their modified from the above sources are shown below in Figure 3.

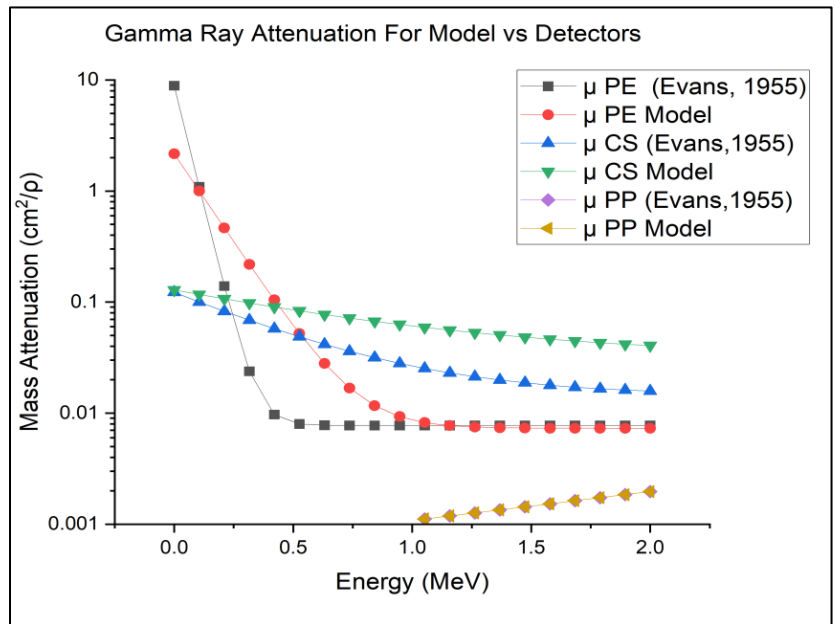

**Figure 3.** Comparison of NaI empirical relations between Knoll's textbook and model. (Knoll, 2010) (Evans, 1955)

Compton scattering remains the most difficult to model because it is the most dominate effect for the energies used in the model and the additional requirements to calculate the scattering angle. This is partially due to the random nature of Compton scattering. While it is relatively easy to determine the energy following a scatter, it is much more difficult to predict the angle for scattering. Not only are the scattering angles, dependent upon the energy of the incoming gamma, but there is a bias for forward scattering with higher energies and backscattering at lower energies. The energy of the gamma following scattering can be found from equation (5) and the bias is based on the *Klein-Nishina* equation (6) (Klein & Nishina, 1929):

$$\lambda' - \lambda = \frac{h}{m_0 c}(1 - cos(\theta)) \tag{5}$$

$$\frac{\delta\sigma}{\delta\Omega} = \frac{1}{2}r_e^2 (\frac{\lambda}{\lambda'})(\frac{\lambda}{\lambda'}) + (\frac{\lambda'}{\lambda}) - sin^2(\theta) \tag{6}$$

Where λ is the initial gamma wavelength, λ' is the scattered gamma wavelength, h is the plank's constant, $m_0c^2$ is the rest mass of an electron, $\theta$ is the scattering angle, $r_e$ is the classical electron radius, and $\frac{\delta\sigma}{\delta\Omega}$ is the angular distribution of scattered gammas. The sample angles for scattering are generated during the initialization of the hybrid Monte-Carlo and based on the *Klein-Nishina* for an initial energy of 0.5 MeV. Figure 4 shows this distribution and the probability of occurrence.

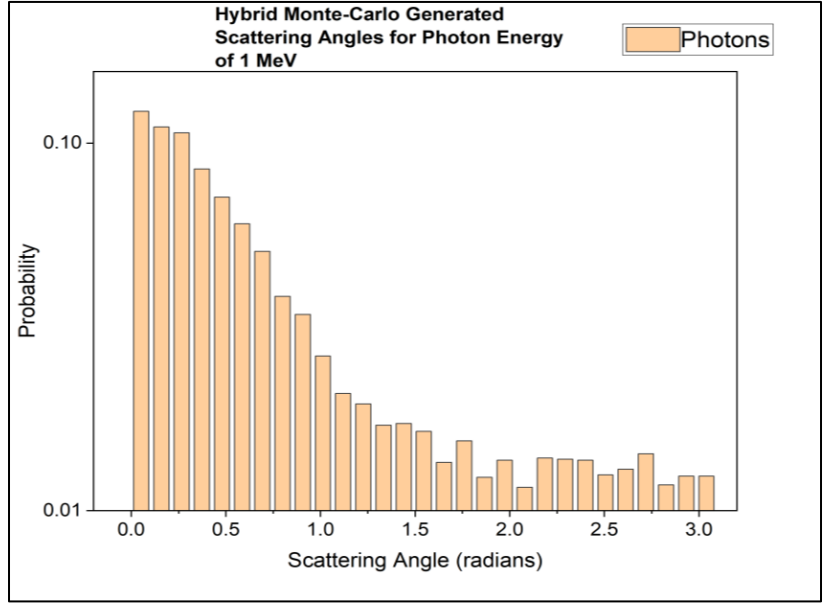

**Figure 4.** Distribution of scattering angles based on *Klein-Nishina* for 0.5 MeV.

Pair-production is a binary response in the MATLAB script, where the energy of the current gamma is compared to 1.02 MeV. If the current energy is greater, than 1.02 MeV, a positron-electron production and subsequent positron annihilation occurs if a variable set to random number between 0 and 1 is greater than another random number of the same characteristic. The relationship is shown in equations (7) and (8). Both values are produced through the Mersenne Twister method as MATLAB's standard random number generator. (Matsumoto & Nishimura, 1998) A single-escape or double escape event is then simulated by another variable defined by MATLAB's standard random number generator.

$$\gamma \rightarrow e^+ + e^- \tag{7}$$

$$e^+ + e^- \rightarrow 2\gamma \tag{8}$$

The photoelectric effect is simulated by a creating a photopeak based on user inputted resolution at the initial gamma energies that each radioactive source emits, e.g., Co-60 at 1.17 and 1.33 MeV. The resolution is calculated by assuming that the resolution for a NaI at Cs-137 is 8% and HPGe is 0.8%. Using the NaI the resolution was established by using FWHM of the three isotopes. Then a linear fit was applied to produce equation (9). The resolution requires a user input for a detector of a higher resolution than NaI. The correction factor ranges from 0.1 – 1 and corresponds to a user input of 100% and 0% respectively.

$$Resolution = Correction\ Factor * (-0.0298 * Energy + 0.08972) \tag{9}$$

While the resolution can vary for each source, using Cs-137 where the peak gamma is near the center of the ranges displayed for these models, the assumption is that the resolution is subjectively sufficient for all gammas between 0.5 MeV - 2 MeV. It is important to note that the user specified input when at 0% is actually 0.01% of the correction factor. This prevents a null resolution value, and is a result of MATLAB requiring the input to be between 0-100. When a user selects 0% the model simulates a NaI detector and 100% results in a HPGe detector. Figure 5 shows the linear estimated resolution for a 0% user input compared to the benchmarked data used to derive equation (9).

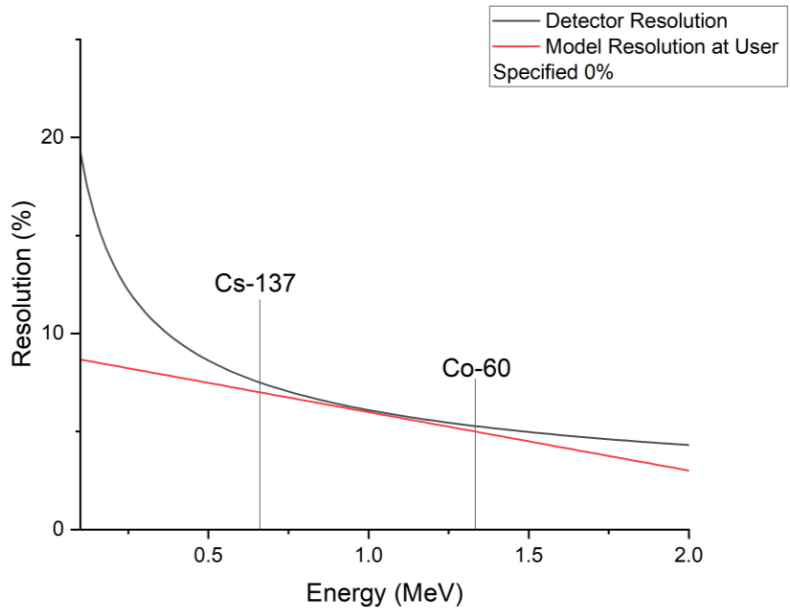

**Figure 5.** Comparison of resolution for benchmarking detector and model with user input 0% corresponding to a correction factor of 1 in equation (9).

### 3.) Benchmarking

The benchmarking of both simulators was conducted with NaI and HPGe spectra of approximately 1-microcurie each of Co-60, Cs-137, and Na-22 at 100 seconds, 300 seconds and 600 seconds with 1024 channels/795 Volts/2.4 Gain for NaI and 16384 channels/2000 Volts/2.4 Gain for HPGe. The spectra collected for benchmarking the simulation required a collimated beam of gammas to ensure even energy deposition and the loss of gammas between the source and scintillating material. An uncollimated beam of gammas allows the gammas to spread isometrically. To prevent this, spectra were taken through a lead collimator on a 3-D printed support base as seen in figure 6. The collimator is a 2-inch x 4-inch x 4-inch block of lead with a 0.25-inch diameter hole bored through the entirety and a 1-inch diameter hole is centered on the smaller and bored 1 inch deep to hold the source. A similar approach was taken for the HPGe detector, however, due to the orientation of the detector, a longer neck was required on the 3D printed source/collimator housing. It is critical to note that the experimental setup for both detector types took place within a lead well to minimize the detection of background radiation in the benchmark spectra.

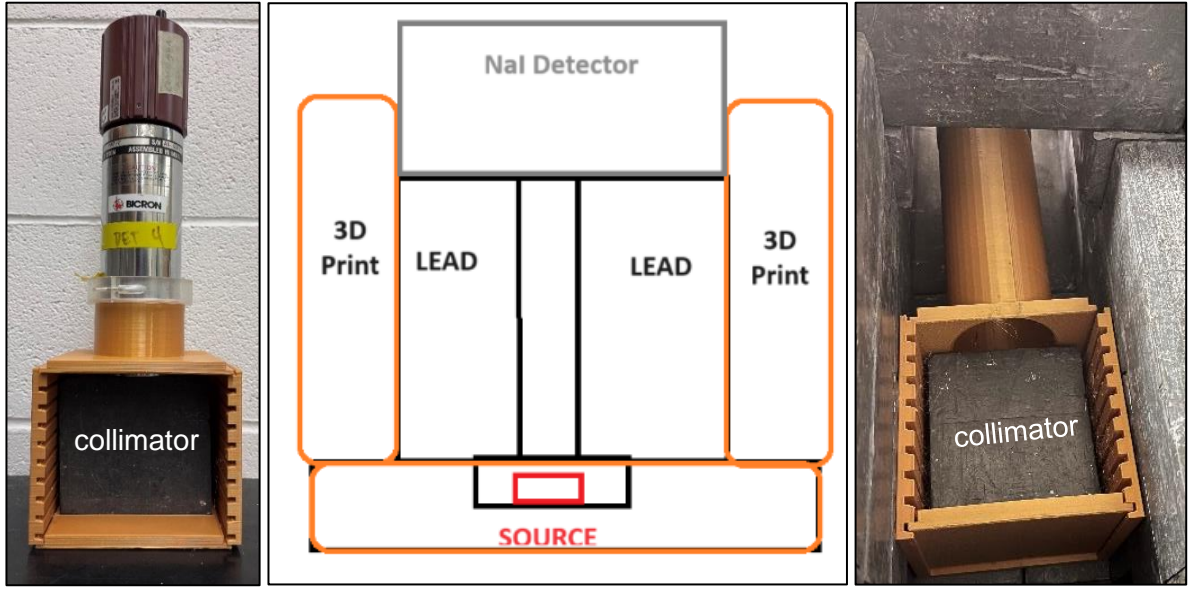

**Figure 6.** Experimental setup for the NaI (Left) and HPGe (Right) detector benchmarking. The setup placed in a lead well to minimize background.

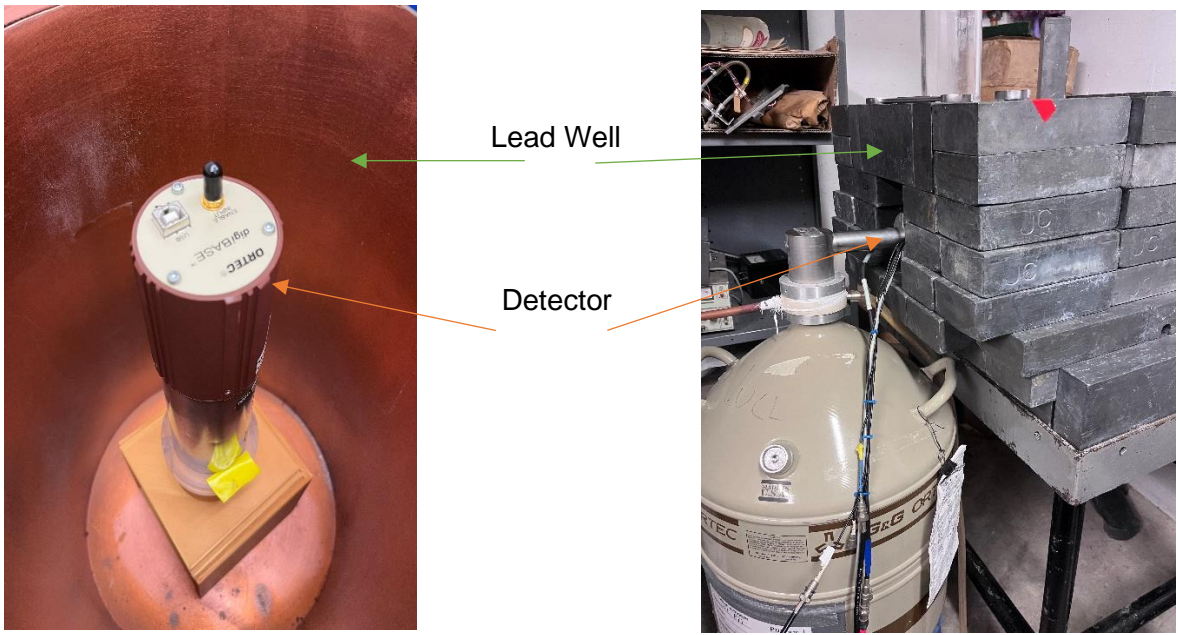

**Figure 7.** Experimental setup for the NaI (Left) and HPGe (Right) while placed in their respective lead wells to minimize background radiation.

Figure 8 compares the difference between the simulation and a 600 second benchmark spectrum for a 20-year-old, 1-microcurie sample of Co-60. Figure 9 provides relative error between the experimental results and the Purdue model. The upper panel displays the overall relative error for all channels, the middle panel shows channels relevant to the Compton continuum, and the lower panel displays channels relevant to the photo peaks.

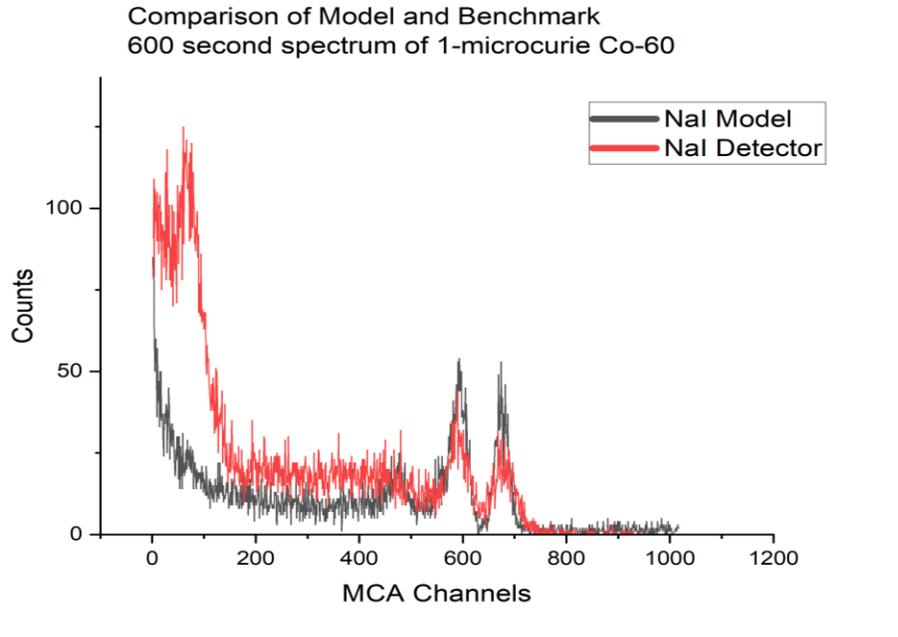

**Figure 8.** Comparison of NaI Co-60 simulation in GUI for "600 seconds" with experimental data.

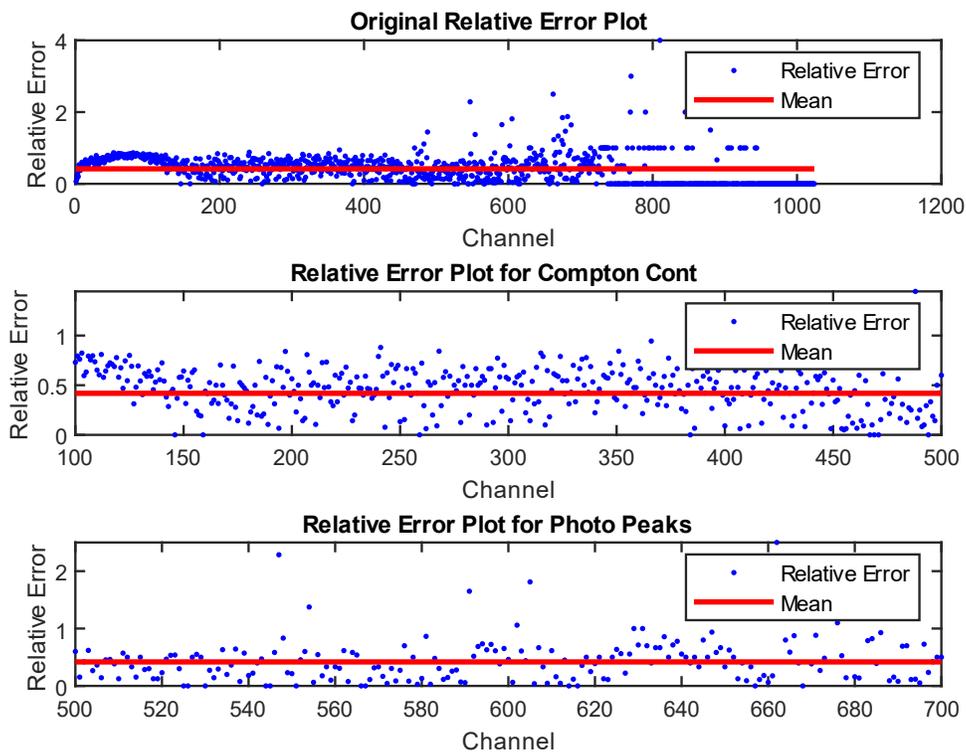

**Figure 9.** Relative error between NaI Co-60 simulation for "600 seconds" with experimental data.

At the surface, the model overlaps experimental data when considering the main architecture of the spectra. The Compton continuum and the photo peaks are nearly aligned over when viewing all channels. There is some variation in the counts between the model and experimental data indicating that the total number of photons tracked is not fully accurate for the activity of the actual Co-60 sample. This is to be expected, as radioactive decay is a trend and not absolute over time. In future editions of the code, this can be corrected by modifying the time constants for decay.

Considering the relative error between the data points, there is more error in the low and high channels. In the channels under 150, this error can be accounted for the lack of consideration if background radiation and x-rays in the code. This is again apparent in figure 8 where these features are present in the experimental data, but not the model. The higher channels experience much larger relative error, with values of up to 400%. This occurrence can be explained by the low counts at higher energy in the experimental data, and the noise generated in the code. For example, if the NaI detector receives zero counts in a particular channel, then the model randomly generates four counts in the same channel to represent noise, the high relative error will be present. A different mode of presenting error such as root mean square, may suppress the extremes. However, using the simple relative error shows the important features such as the Compton Continuum and photo peaks in an appropriate manner when comparing the data spectra as shown in figure 8. In the lower panels of figure 9, there is a relative error average of about 45%. For two signals with random noise, 45% relative error is fair. When the comparison between two spectra of experimental data can produce relative errors on the same order of magnitude.

A similar comparison can be made for the HPGe model setting and experimental data. Figure 10 compares the difference between the simulation and a 100 second benchmark spectrum for a 20-year-old, 1-microcurie sample of Co-60. Figure 11 shows the relative error between the model and experimental data

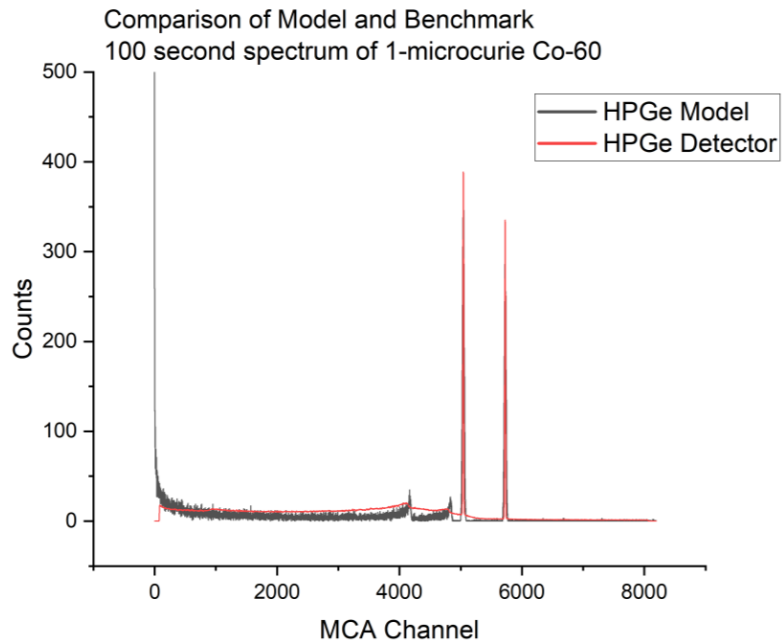

**Figure 10.** Comparison of HPGe Co-60 simulation in GUI for "100 seconds" with experimental data.

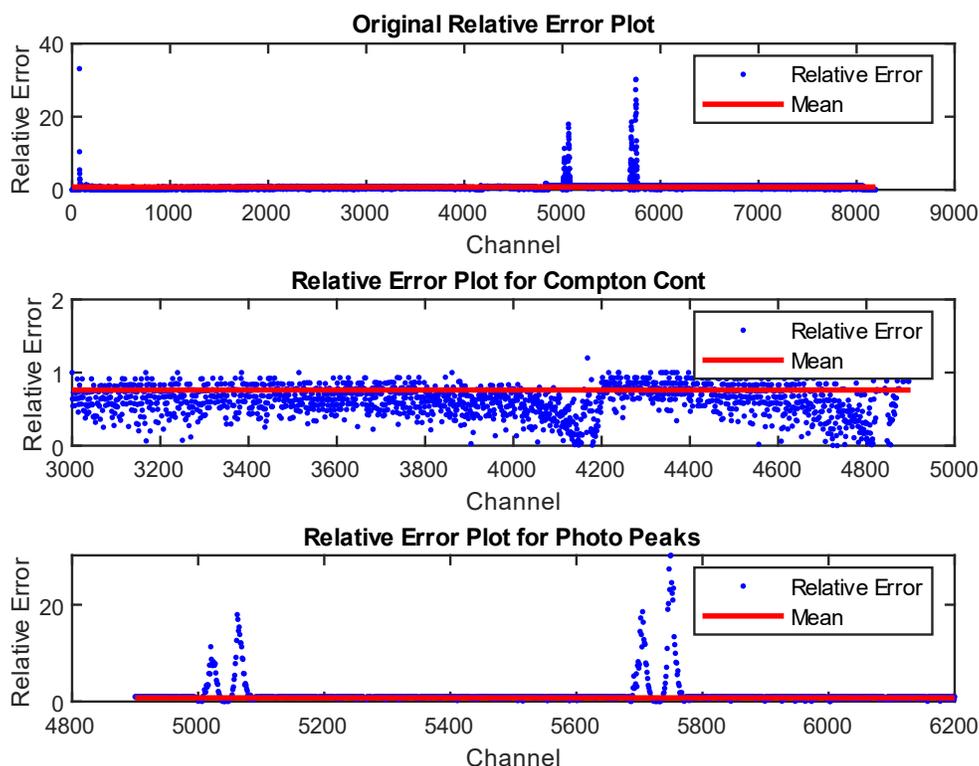

**Figure 11.** Relative error between HPGe Co-60 simulation for "100 seconds" with experimental data.

Again, considering the relative error between the model and experimental data, the average is approximately 80%. However, there are different features that occur when comparing figure 11 and 9. The data overall has less variance with the exception of the photo peaks. Given that the channels are not exactly aligned and the high resolution of the HPGe detector, any misalignment between the channels of each data set will lead to large error. This is apparent in the last panel of figure 11, where the two data sets do not perfectly line up. Counter to the photo peaks, the significantly larger number of channels in a HPGe detector reduces the variance in the data for the Compton continuum.

Similar to NaI, it is critical to note, there is noise present in both experimental and model data which leads to what would normally be considered high relative error. As a result, a comparison on how the two signals align with their spectra architecture and relative error must be considered together. By doing so, evidence supports that the model in both HPGe and NaI modes align reasonably well with their respective experimental detectors.

**Graphical User Interface**

The Graphical User Interface offers a clean and organized means of adjusting the algorithm's input variables. The user can select a up to 8 pre-loaded 1 microcurie gamma emitting isotopes or create their own between 1 and 2 microcuries with an option of two initial gamma energies. Table 1 provides a list of isotopes loaded into the code. Then they can select the number of MCA channels available. Also included is the ability to select the isotope age in years or days depending on the initial isotope selected. The count time is adjustable from 1 to 3000 seconds. Furthermore, the user can select the detector's resolution from 0 to 100 percent. This allows for the response of several detectors or to display the three-gamma interaction without resolution. Perhaps the most useful feature is the ability to download the spectrum of output in a comma separated value file. Allowing the user to manipulate the values allows for a similar experience and universal utility that occurs with an actual detector. User input variables increases the utility of the model for various applications seen fit by the user.

Table 1. List of Isotopes Pre-Loaded in Purdue Code

| Isotope Selection | |
|---|---|
| Co-60 | Th-228 |
| Co-57 | Am-241 |
| Na-22 | Au-198 |
| Ba-133 | Cs-137 |

   Earlier versions of the algorithm relied on single stream variable computation. This significantly slowed the computation of several combinations of user input variables. The solution was to add parallelization into the script. An example of this, was running 20-year-old Cs-137 for a count time of 600 seconds. Without parallelization the algorithm required 2 minutes and 27 seconds to return a response. The addition of multiple cores reduced that time to just under 3 seconds. However, a consequence of running multiple cores, requires a longer startup time. For newer systems it can take 10 seconds to start multiple cores, and up to 60 seconds for older systems running MATLAB. This is done in the background of the GUI when the power button is pressed, but this "dead time" can be analogous to the application of voltage to a real detector. Figure 12 shows the options available to the end user.

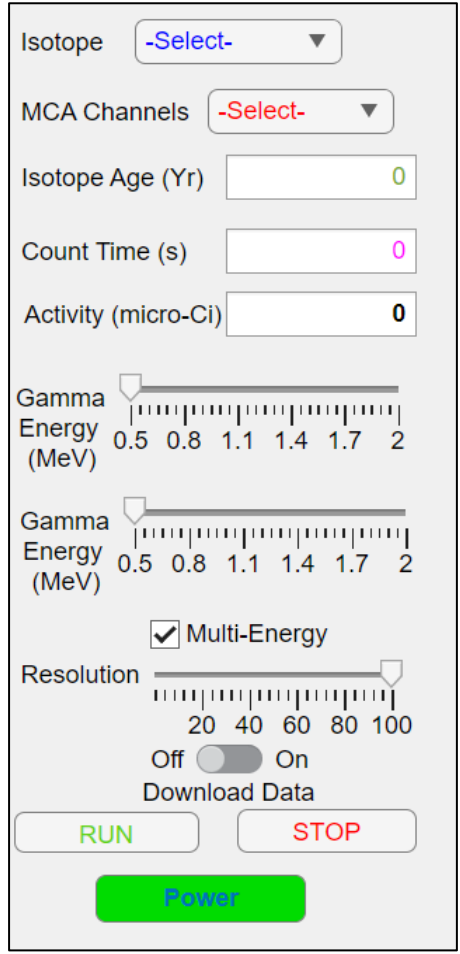

**Figure 12.** Graphics user interface view of user input variables for HPGe (100% resolution).

## 4.) Results

    The goal of the model was to provide flexibility to the user to simulate a wide range of settings that are present on actual detectors. The resolution feature can be seen between different detector brands and different scintillating materials.

    An example of the versatility can be expressed for a 20-year-old 1-microcurie sample of Cs-137 counted for 600 seconds for different resolutions. Figure 13 shows the result at 100% resolution and figure 14 at 20% resolution.

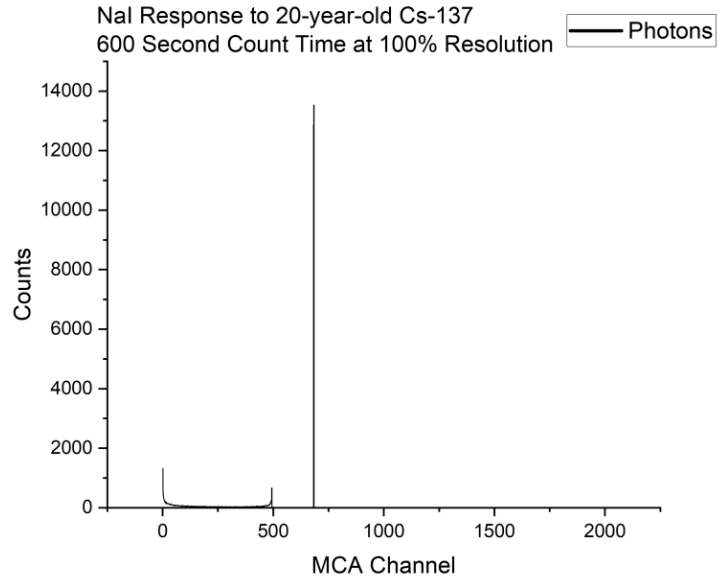

**Figure 13.** Results for Cs-137 at user specified 100% resolution. Used to show HPGe response.

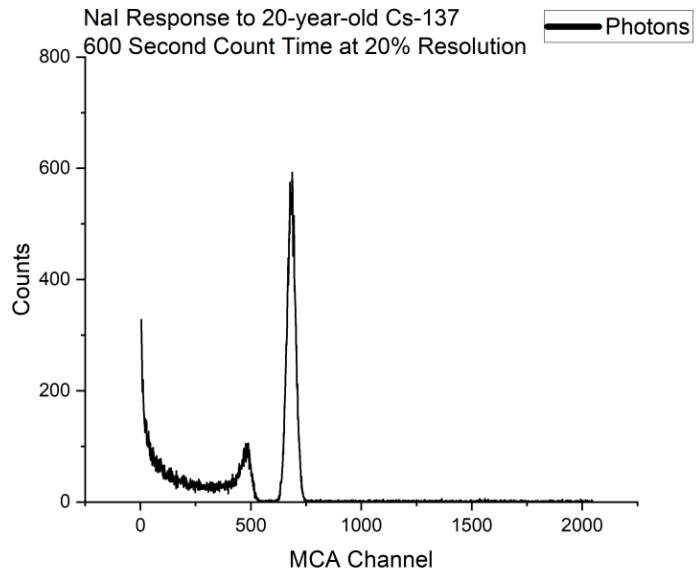

**Figure 14.** Results for Cs-137 at user specified 20% resolution. Used to show NaI response.

Another example expressing the flexibility of the model is the ability to modify the sample age. Take Ba-133, which has a half-life of 10.51 years. By modifying the sample age of a 1-microcurie sample from 5 years to 50 years, the responses are displayed in figures 15 and 16 respectively.

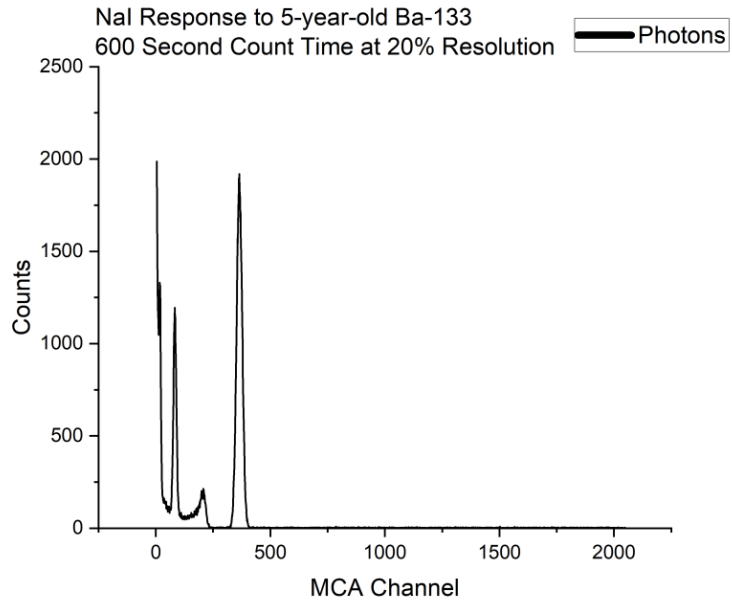

**Figure 15.** Results for Ba-133 at user specified 5-year-old sample and 600 second count time.

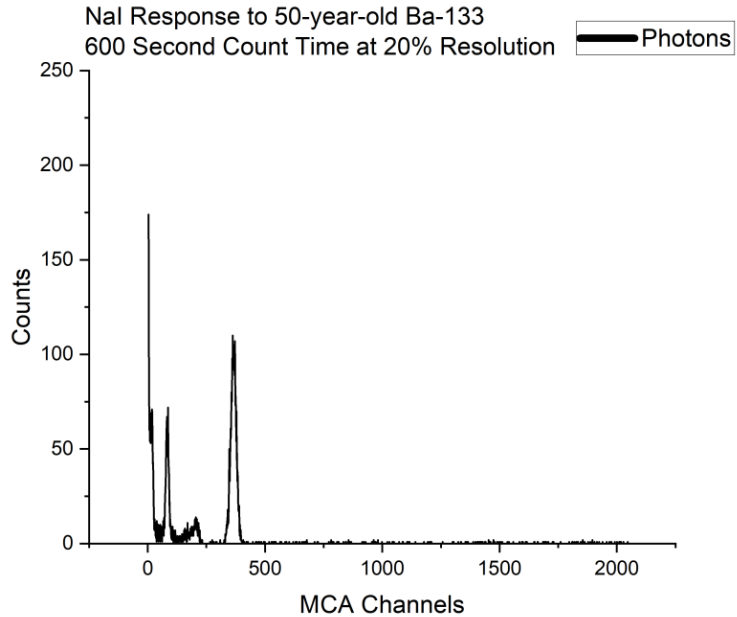

**Figure 16.** Results for Ba-133 at user specified 50-year-old sample and 600 second count time.

The model is also capable of producing gamma spectrum with a combination of up to three isotopes and an option to include background radiation. Background is difficult to produce because of the large variety of naturally occurring isotopes for given locations, building types, or even altitudes from ground level. However, a basic spectrum representing the background counts of from U-235 and Ra-226 decay chains as well as naturally occurring K-40, Cu-63, and Cu-65 is present. The model assumes a 0.25 µCi activity of all the isotopes with a half-life of 1000 years. While none of these values are truly accurate, the background spectrum produced is consistent, with many background spectra. (Arnold, Heckel, & Wershofen, 2019) Figure 17 shows a combination spectrum of 1-µCi each of Ba-133, Co-60, and Na-22 at an age of 1-year and a count time of 600 seconds. Conversely, figure 18 displays the same combination at 100 years, which is well beyond 5-half lives of all the isotopes. The result shown in figure 18 is the simulated background spectrum. The user interface allows for the inclusion of the background counts with a selection of a checkbox. For most applications, the background is much lower than the counts and does not affect the results of the spectrum, but it is evident that it plays a role when all the isotopes have passed 5-half lives and have decayed to nearly null activity.

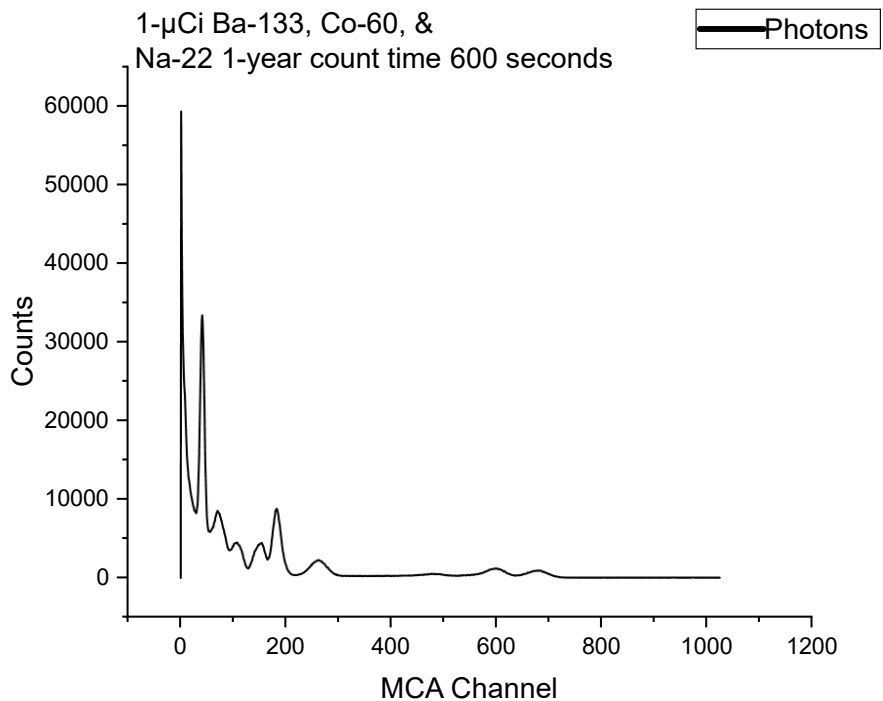

**Figure 17.** NaI response for 1-µCi of Ba-133, Co-60, and Na-22 at 1-year age and a count time of 600 seconds.

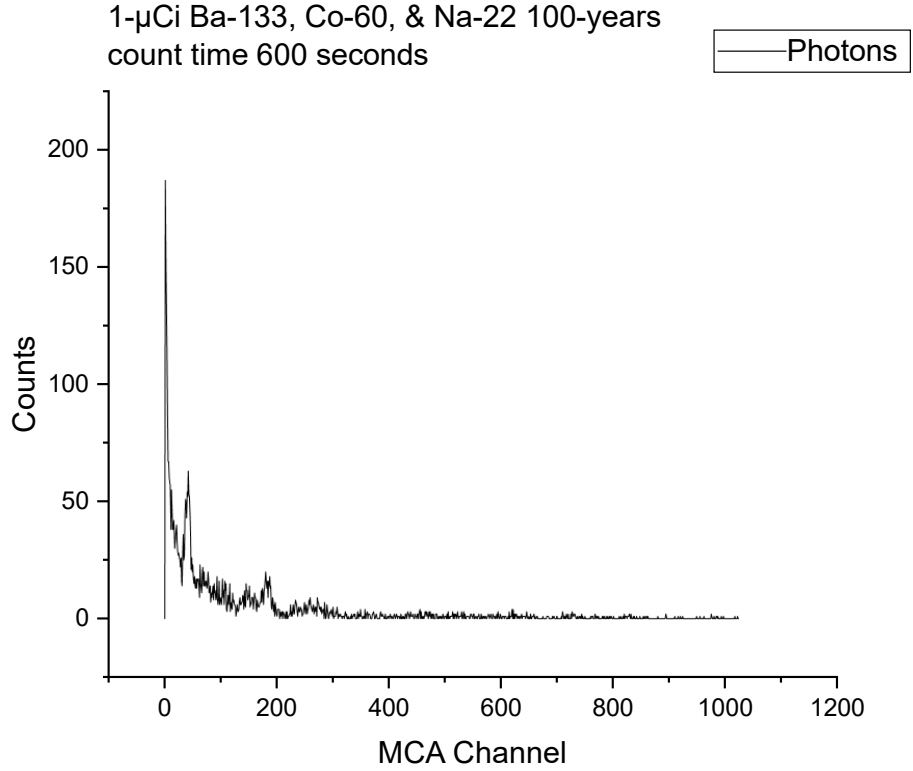

**Figure 18**. NaI response for 1-µCi of Ba-133, Co-60, and Na-22 at 100-years age and a count time of 600 seconds.

The advantage of making use of the hybrid Monte-Carlo methods is the reduced computational times required to conduct calculations. The model was run on two systems and timed to highlight the low computational times. The first system is a Dell OptiPlex with an i7 14700 vPro processor and 16 computational cores. The other was HP EliteBook running an i7-8550U with 4 computational cores. The results of different combinations of the model running for Co-60 are displayed in table 2. Different combinations of isotopes are shown in table 3.

Table 2. Computation time in seconds for variations of Co-60 between two processors.

|  | Dell i7 14700 vPro | HP i7-8550U |
|---|---|---|
| 1 µCi/1 Year/100 seconds/256 Channels Co-60 | 0.47466 | 1.64626 |
| 1 µCi/1 Year/100 seconds/2048 Channels Co-60 | 0.485047 | 2.095934 |
| 1 µCi/1 Year/1 second/2048 Channels Co-60 | 0.244663 | 0.827295 |
| 1 µCi/1 Year/3000 seconds/2048 Channels Co-60 | 5.304477 | 15.77539 |
| 1 µCi/0.1 Years/100 seconds/2048 Channels Co-60 | 0.353323 | 1.400918 |
| 1 µCi/100 Years/100 seconds/2048 Channels Co-60 | 0.134576 | 0.767599 |
| 0.1 µCi/1 Year/100 seconds/2048 Channels Co-60 | 0.102322 | 1.072231 |
| 10 µCi/1 Year/100 seconds/2048 Channels Co-60 | 1.042232 | 6.820869 |

Table 3. Computation time in seconds for variations of isotopes between two processors.

|  | Dell i7 14700 vPro | HP i7-8550U |
|---|---|---|
| 1 µCi/1 Year/100 Seconds/2048 Channels Co-60 | 0.288089 | 1.437798 |
| 1 µCi/1 Year/100 Seconds/2048 Channels Cs-137 | 0.34729 | 1.593292 |
| 1 µCi/1 Year/100 Seconds/2048 Channels Na-22 | 0.131442 | 1.009565 |
| 1 µCi/1 Day/100 Seconds/2048 Channels Au-198 | 0.127421 | 0.944222 |
| 1 µCi/1 Year/100 Seconds/2048 Channels Th-228 | 0.166742 | 1.339904 |
| 1 µCi/1 Year/100 Seconds/2048 Channels Co-60, Cs-137 | 0.347928 | 2.203138 |
| 1 µCi/1 Year/100 Seconds/2048 Channels Co-60, Cs-137, Na-22 | 0.405756 | 2.93693 |

**5.) Comparison with GADRAS**

        One of the Oakridge National Lab Codes that attempts to solve the detector response is Gamma Detector Response and Analysis (GADRAS). (Mitchell & Mattingly, 2009) Similar to MCNP, GADRAS requires a vetting process which often takes months and has strict export control parameters. Additionally, the popularity of the software is much less than MCNP, and has sparse 3rd-party documentation. As a result, the user manual is the only means to understanding the software.

        Objectively, GADRAS is much more robust than the Purdue model. The user is allowed to create unique detectors with specific parameters, the background radiation of multiple locations globally can be considered, and there is a large library of sources to include neutrons. However, GADRAS is an analytical solution that produces a solution that does not vary between runs and does not replicate detector noise. The Purdue model has an energy resolution option, which can theoretically be set to match any scintillating material provided in the detector selection and parameter page. At this point in development, the user can only adjust the size of the scintillating crystal in the Purdue code. Another limiting difference in the Purdue model is the small number of isotopes. Currently, there is an option to combine up to 3 of the isotopes combined at any given run which effectively provides 336 combinations of isotopes.

        When comparing the two models for a given isotope, there are a few critical differences. GADRAS allows for the energy calibration of the detector, while the Purdue code does not. As a result, the channels which the energy peaks are displayed are distorted. The Purdue model is slightly expanded when compared to GADRAS. This effect is a direct effect of the bias voltage used when samples were collected for benchmarking and built into the code. GADRAS does not have the ability to vary counting time, and defaults to 10 seconds of live time. To make a useful comparison between GADRAS and the Purdue code using a 3"x3" NaI crystal, the Purdue code was set to the following: 1 µCi Co-60, 1- year old sample, counting time of 10 seconds. The raw results are shown in figure 19, and the results to highlight the difference in spectrum compression are displayed in figure 20 by aligning the 1.17 MeV peaks. This was done by finding the channel with the peak in the GADRAS code, then artificially setting the same peak of the Purdue code at that channel and adjusting the rest accordingly. Due to this, the Purdue code has negative channel numbers, but the absolute value of the channels remains 512.

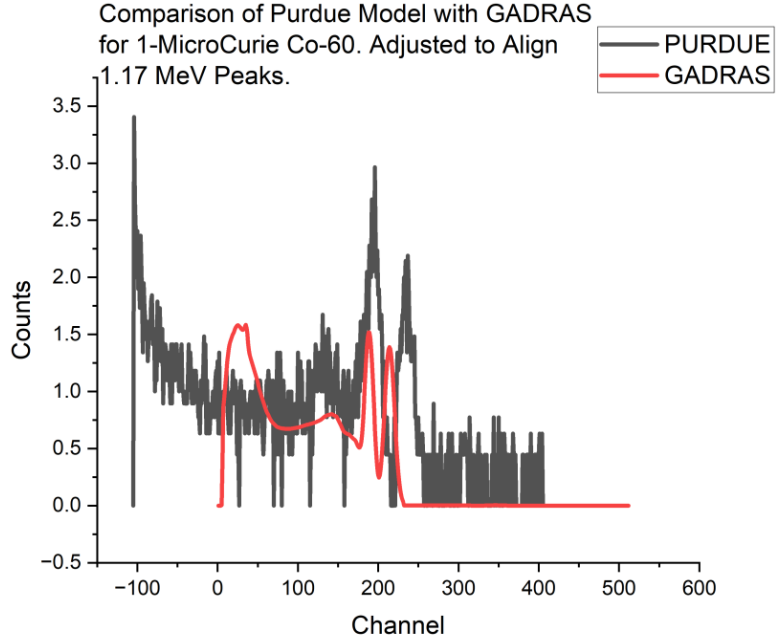

**Figure 19.** GADRAS and Purdue code comparison for 1 µCi Co-60, 1-year old sample, and 10 second count time.

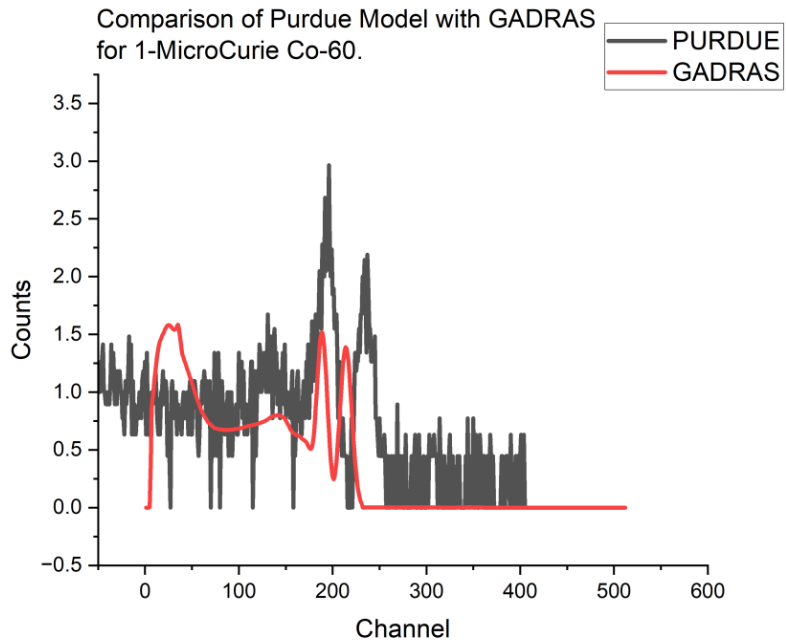

**Figure 20.** GADRAS and Purdue code comparison for 1 µCi Co-60, 1-year old sample, and 10 second count time. Adjusted to align 1.17 MeV peaks.

To further highlight the similarities between the two codes outputs, the standard HPGe detector was setup in GADRAS for the same Co-60 sample parameters listed previously. However, the energy resolution in the Purdue code was set to 100%. Figure 21 shows the raw output of both files and figure 22 displays an adjusted channels aligned at the 1.17 MeV peaks. The same apparent data compression is apparent with the energy resolution showing that channel number is not a function of resolution in this code.

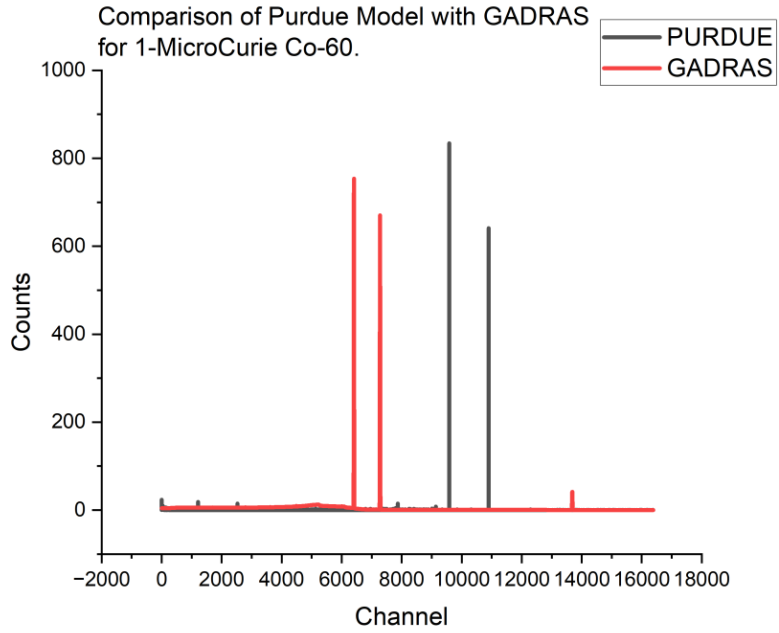

**Figure 21.** GADRAS and Purdue code HPGe comparison for 1 µCi Co-60, 1-year old sample, and 10 second count time.

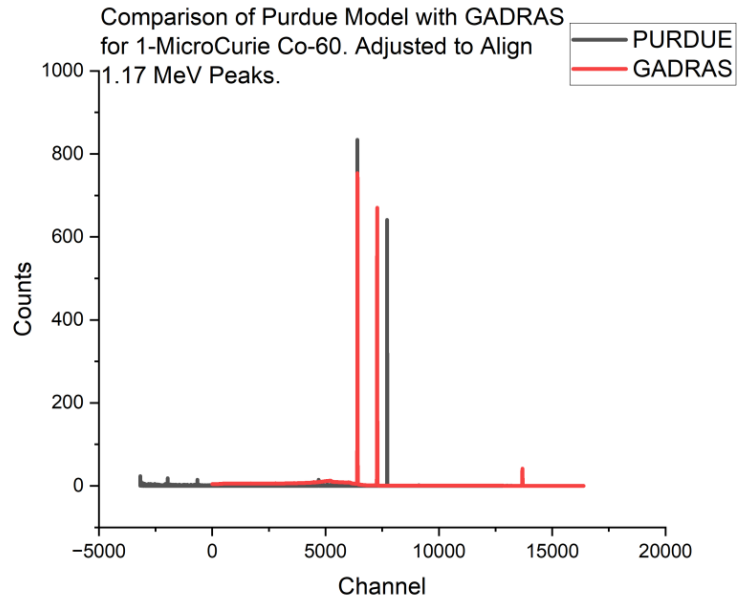

**Figure 22.** GADRAS and Purdue code HPGe comparison for 1 µCi Co-60, 1-year old sample, and 10 second count time.

An additional comparison that can be made between the codes is the performance with multiple isotopes. It should be noted that GADRAS does not combine the spectrum of each isotope on the displayed plot. Rather, this was conducted in a spreadsheet by summing the output of each isotope from the CSV file. The benefit of the Purdue code is this process is done automatically in the GUI. To highlight the similarities, samples of Co-60, Cs-137, and Na-22 at 1 year-old and 1 µCi were counted at 10 seconds for the standard GADRAS NaI detector. The results are shown below in figure 23. Similar to before, figure 24 shows an alignment of one select peak for better reference. The Cs-137 662 keV peaks are aligned.

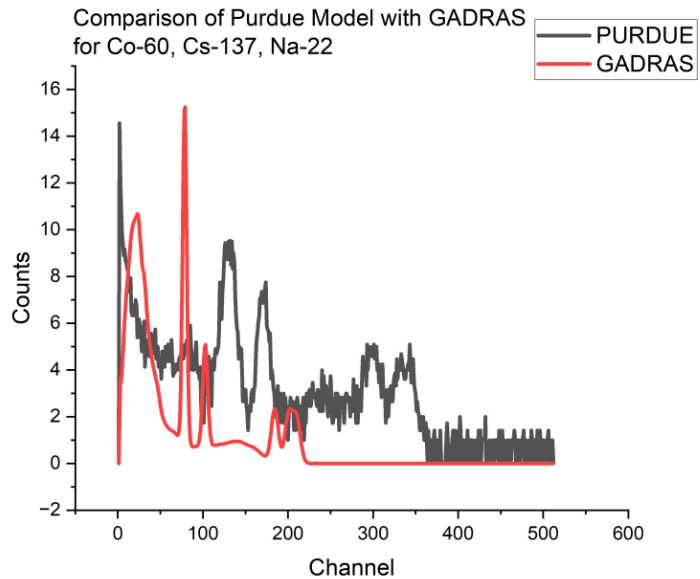

**Figure 23.** GADRAS and Purdue code HPGe comparison for 1 µCi Co-60, Cs-137, Na-22, 1-year old samples, and 10 second count time.

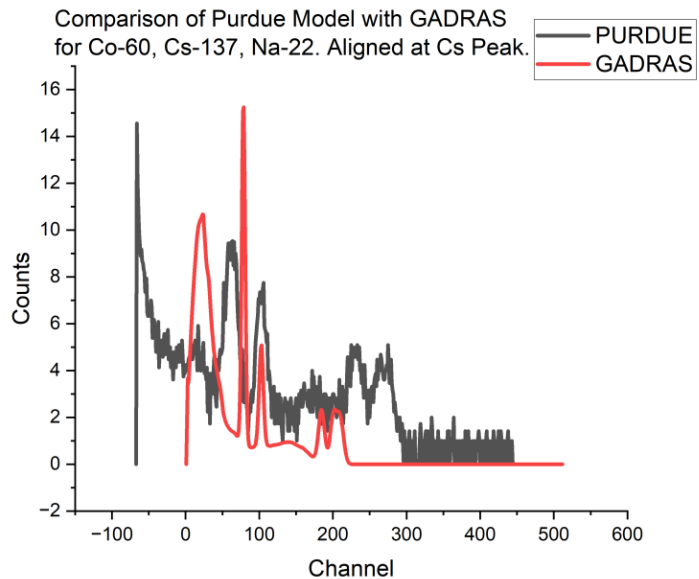

**Figure 24.** GADRAS and Purdue code HPGe comparison for 1 µCi Co-60, Cs-137, Na-22, 1-year old samples, and 10 second count time. Adjusted to be aligned at 662 keV.

## 6.) Conclusion

While an algorithm to fully encapsulate the physics that occur within a gamma detector would be nearly impossible to build, MCNP is currently the industry standard for producing models for these complex interactions. The problem with MCNP is the software requires large computing requirements, an understanding of the language, and a clearance to use the software. For an educational setting, these requirements prevent a vast majority of people from learning about the detection of radiation.

Other codes specifically developed for the ability to model gamma detectors, such as GADRAS are a great analytical solution. However, like its MCNP counterpart, it is difficult to gain a license. Additionally, the software lacks the ability to produce the noise and random variation in the detector response that a numerical code may produce. When comparing the Purdue code and GADRAS side by side, there is evidence to support that the results for a several combinations of inputs produces similar results.

The function of the Purdue code allows users to gain an understanding of radiation and the underlying physics without the requirement of having access to expensive equipment or MCNP/GADRAS. The use of hybrid Monte-Carlo methods decreases the computation requirements. A code that could take minutes to hours to run in traditional Monte-Carlo codes can take a matter of seconds or less in the Purdue code. This is dependent upon the type of processor used, but the correlation between specific tasks on different machines produces a time change that is proportional between the two processors.

Future work for the Purdue code includes the addition of a robust isotope library which can match that of GADRAS. Currently, a user can apply up to three isotope combinations of the pre-loaded 8 isotope library, but this can be expanded to allow for more complex spectra to be created. Additionally, the inclusion of calibration or bias voltage-like feature would be helpful for creating spectra that can be are expanded to match the calibration of any detector. Furthermore, the inclusion of neutrons *via* gas-filled detectors would make the Purdue code a more competitive model. Although, there is no claim that the Purdue code can replace widely used transport codes. It may offer an alternative that can be used in an educational setting by more people, with comparable results to experimental data and data produced by industry leading gamma detector response codes.

**Acknowledgments**

This research was performed using funding from the Purdue College of Engineering and Purdue Military Research Institute.

**Disclaimer**

*The views expressed in this article are those of the author and do not reflect the official policy or position of the United States Air Force, Department of Defense, or the U.S. Government.*

**Code Distribution**

The MATLAB file for GammaGenius can be found on MATLAB File Share at the url: https://www.mathworks.com/matlabcentral/fileexchange/167811-gammagenius

# Appendix A: Program Pseudo Code

```
FUNCTION main():
  CONSTANTS:
    PI = 3.14159
    CF = 2.92e20
    re = 2.818e-15
    eV_to_J = 1.60218e-19

  VARIABLES:
    Isotope1 = "Co"
    Isotope2 = "Na"
    Isotope3 = "Cs"
    time = 300
    T1 = 20
    T2 = 20
    T3 = 20
    E = [1.33, 1.17]
    E2 = [0.511, 1.278]
    Ac = 1
    Det_Resolution = 'Y'
    MCA_channels = 1024
    Use_res = 1
    iteration = 3
    D = 7.62
    IsotopeStorage[iteration]
    TimeStorage[iteration]

  FOR i = 1 TO iteration:
    IsotopeStorage[i] = GET_Isotope(i)
    TimeStorage[i] = GET_Time(i)

  FOR p = 1 TO iteration:
    Isotope = IsotopeStorage[p]
    T = TimeStorage[p]
    activity = CALC_Activity(Isotope, T)
    CALC_Parameters(Isotope, T)
    gamma_rate, photons = CALC_Gamma_Rate(activity, time)
    CALC_Cross_Sections(Isotope)
    CALC_Interaction_Probability(Isotope)
    Edep = CALC_Energy_Deposition(Isotope, gamma_rate, photons)
    STORE_Edep(Edep, p)

  histdata = CONCATENATE_Histograms(iteration)
  csv_data_mat = CREATE_CSV_Data(histdata, MCA_channels)
  SAVE_CSV_File(csv_data_mat)

FUNCTION GET_Isotope(index):
  SWITCH index:
    CASE 1:
      RETURN "Co"
    CASE 2:
      RETURN "Na"
    CASE 3:
      RETURN "Cs"

FUNCTION GET_Time(index):
  SWITCH index:
    CASE 1:
      RETURN 20
    CASE 2:
      RETURN 20
    CASE 3:
      RETURN 20

FUNCTION CALC_Activity(Isotope, T):
  SWITCH Isotope:
    CASE "Co":
      lam = 0.693 / 5.27
      N_init = 2.22e15
      RETURN (lam * N_init * EXP(-lam * T)) / CF
    CASE "Na":
      lam = 0.693 / 2.6
      N_init = 1.1e15
      RETURN (lam * N_init * EXP(-lam * T)) / CF
    CASE "Cs":
      lam = 0.693 / 30
      N_init = 1.26e16
      RETURN (lam * N_init * EXP(-lam * T)) / CF

FUNCTION CALC_Parameters(Isotope, T):
  IF Isotope == "Co":
    E_init = [1.33, 1.17]
  ELSE IF Isotope == "Na":
    E_init = [0.511, 1.278]
  ELSE IF Isotope == "Cs":
    E_init = 0.667
  ELSE:
    E_init = [0.511, 0.4118]

  // Calculate other parameters based on Isotope and T

FUNCTION CALC_Gamma_Rate(activity, time):
  gamma_rate = activity * 3.7e10 * 0.5
  photons = ROUND(time * gamma_rate)
  RETURN gamma_rate, photons

FUNCTION CALC_Cross_Sections(Isotope):
  // Calculate cross-sections based on Isotope

FUNCTION CALC_Interaction_Probability(Isotope):
  // Calculate interaction probability based on Isotope

FUNCTION CALC_Energy_Deposition(Isotope, gamma_rate, photons):
  // Calculate energy deposition based on Isotope, gamma_rate, and photons

FUNCTION STORE_Edep(Edep, p):
  // Store energy depositions for each iteration

FUNCTION CONCATENATE_Histograms(iteration):
  // Concatenate histograms for all iterations

FUNCTION CREATE_CSV_Data(histdata, MCA_channels):
  // Create CSV data from histogram data

FUNCTION SAVE_CSV_File(csv_data_mat):
  // Save CSV file with histogram data

main()
```